\documentclass[aps,prd,amssymb,twocolumn,floatfix,showpacs,preprintnumbers
]{revtex4}
\usepackage{graphicx}
\begin{document}
\preprint{U. C. Irvine Technical Report 2002-14}
\title{ Iso-singlet Down Quark Mixing And {\boldmath $CP$} Violation 
Experiments}
\author{Donovan Hawkins and Dennis Silverman}
\affiliation{Department of Physics and Astronomy, \\
University of California, Irvine \\
Irvine, CA 92697-4575 }
\email{djsilver@uci.edu}
\date{\today}

\begin{abstract}
  We confront the new physics models with extra iso-singlet down
  quarks in the new $CP$ violation experimental era with
  $\sin{(2\beta)}$ and $\epsilon'/\epsilon$ measurements, $K^+ \to
  \pi^+ \nu \bar{\nu}$ events, and $x_s$ limits.  The closeness of the
  new experimental results to the standard model theory requires us to
  include full SM amplitudes in the analysis.  In models allowing
  mixing to a new isosinglet down quark, as in E$_6$, flavor changing
  neutral currents are induced that allow a $Z^0$ mediated
  contribution to $B-\bar B$ mixing and which bring in new phases.  In
  $(\rho,\eta)$, $(x_s,\sin{(\gamma)})$, and $(x_s, \sin{(2\phi_s)})$
  plots we still find much larger regions in the four down quark model
  than in the SM, reaching down to $\eta \approx 0$, $0 \leq
  \sin{(\gamma)} \leq 1$, $-.75 \leq \sin{(2\alpha)} \leq 0.15$, and
  $\sin{(2\phi_s)}$ down to zero, all at 1$\sigma$.  We elucidate the
  nature of the cancellation in an order $\lambda^5$ four down quark
  mixing matrix element which satisfies the experiments and reduces
  the number of independent angles and phases.  We also evaluate tests
  of unitarity for the $3\times3$ CKM submatrix.
\end{abstract}
\maketitle

\section{Introduction}
The ``new physics'' class of models we use are those with extra
iso-singlet down quarks, where we take only one new down quark as
mixing significantly.  An example is E$_6$, where there are two down
quarks for each generation with only one up quark, and of which we
assume only one new iso-singlet down quark mixes strongly.  This model
has shown large possible effects in $B-\bar B$ mixing
phases \cite{Silverman:1998uj}.  The new $B$ factory results on
$\sin{(2\beta)}$ in the SM range, the $\epsilon'/\epsilon$
experimental convergence, the new $K^+ \to \pi^+ \nu \bar{\nu}$
result, the $\Delta m_s$ limits near the SM prediction, and other new
measurements require a finer analysis and a potential challenge to new
physics models.  In this paper we include the full SM contributions as
well as the new physics contributions from the iso-singlet down quark
model to jointly analyze the constraints from all of these
experiments, as well as other flavor changing neutral current (FCNC)
limits and SM CKM matrix element constraints.

In models allowing mixing to a new iso-singlet down quark (as in
E$_6$) flavor changing neutral currents are induced that allow a $Z^0$
mediated contribution to $B-\bar B$ mixing and which bring in new
phases \cite{Choong:1994gq,Silverman:1996rk, Silverman:1998uj}.  In
$(\rho,\eta)$, $(x_s,\sin{(\gamma)})$, and $(x_s, \sin{(2\phi_s)})$
plots we still find much larger regions than in the SM, reaching down
to $\eta \approx 0$, $0 \leq \sin{(\gamma)} \leq 1$, and
$\sin{(2\phi_s)}$ down to zero (below the SM range), all at 1$\sigma$
limits.  The nature of the cancellation in a fourth down quark matrix
element $V_{4d}$ to satisfy the experiments is elucidated.  We also
establish ranges for the new mixing elements to the new iso-singlet
down quark, and make a simple estimate of the lower mass limit of the
new down quark.

In Section II we introduce the scenario with more down quarks as in
E$_6$, truncate it to one extra down quark, introduce the $4\times4$
mixing matrix, and apply it to $B-\bar{B}$ mixing. Section III
presents the $CP$ violating $B_d$ and $B_s$ decay asymmetries, and $B_s$
mixing, including the FCNC tree diagram additions.  Section IV
presents the full SM contributions as well as the four down quark
model (FDQM) amplitudes for the $CP$ violating and FCNC $K$ meson
experiments that are used.  Section V presents the joint chi-squared
analysis and results for the SM and FDQM model for the various plots
listed above.  Section VI presents the sizes or limits on the matrix
elements, mixing angles, phases, FCNC couplings and unitarity
quadrangles.  Section VII lists the conclusions and
projects what the next down quark mass limit might be.

\section{Iso-singlet Down Quark Mixing Model}
Groups such as E$_6$ with extra SU(2)$_L$ singlet down
quarks \cite{Hewett:1989xc} give rise to flavor changing neutral currents
(FCNC) through the mixing of four or more down
quarks \cite{Silverman:1996rk,
  Shin:1989eu,Nir:1990hj,Nir:1990yq,Silverman:1992fi}.  The initial
quarks of definite weak isospin in E$_6$, for each generation are: the
left handed iso-doublet $(u^0_{iL}, d^0_{iL})$, their right handed
iso-singlets $u^0_{iR}$ and $d^0_{iR}$, and the yet to be found
iso-singlet pairs $D^0_{iL}$ and $D^0_{iR}$.

We can take the initial up quark matrix to be the mass eigenstates, so
$u^0_i = u_i$, giving $V^u = I_{3\times3}$.  The down quarks $( d^0_i,
D^0_i)$, which correspond to the same generations as $u_i$, mix to form
mass eigenstates $(d_i, D_i)$ via the matrix $V^d(6\times6)$, where
$d_{iL}^0 = V^d_{ij} d_{jL}$.  The weak interaction charged current
matrix is then $U = V^{u\dagger} \times V^d$, the $3\times6$ matrix
that is the upper three rows of $V^d$.  The lower three rows of $V^d$
are the three linear combinations of $(d_i, D_i)$ that are the
iso-singlet $D^0_i$ which cannot couple to up quarks by the weak
interactions.

We truncate the $V^d$ matrix to the $4\times4$ matrix using only the
$D$ quark that mixes most (and dropping the superscript $d$ on $V^d$),
giving the Four Down Quark Model (FDQM).  Calling the new down quark
mixture $D$, the weak charged currents of $D$ to $u$, $c$, and $t$
quarks are $V_{tD} = s_{34}$, $V_{cD} = s_{24} e^{-i\delta_{24} }$,
and $V_{uD} = s_{14} e^{-i \delta_{14} }$, which are in the fourth
column.  The fourth row gives the linear combination that is the
initial iso-singlet $D^0_L$.  The complete $4 \times 4$ mixing matrix
was given previously \cite{Choong:1994rb,Botella:1986gb}.  The leading
terms in the $4 \times 4$ down quark mixing matrix with 6 angles and 3
phases are
\begin{widetext}
\begin{equation}
V = \bordermatrix{  & d & s & b & D \cr
u & c_{12}c_{34} & \quad s_{12}c_{34} & \quad s_{13}e^{-i \delta_{13}}&
\quad s_{14} e^{-i \delta_{14} } \cr
c &-s_{12} &  \quad  1      & \quad  s_{23}   & \quad s_{24} e^{-i
\delta_{24} } \cr
t & (s_{12} s_{23} -s_{13} e^{i \delta_{13} } ) &  \quad -s_{23}    &
\quad  1     & \quad s_{34} \cr  
4 & \quad V_{4d} & \quad V_{4s} & \quad V_{4b} & \quad V_{44} \cr },
\end{equation}
where, to leading order in new angles,
\begin{eqnarray}
V^*_{4d} & =& - s_{14} e^{-i \delta_{14} } +s_{24} e^{-i
\delta_{24} } s_{12} - s_{34} (s_{12} s_{23} - s_{13} e^{-i
\delta_{13} } ), \\
V^*_{4s} & =& - s_{24} e^{-i\delta_{24} } 
- s_{14} e^{-i \delta_{14} } s_{12}  
+s_{34} ( s_{23} + s_{12} s_{13} e^{-i \delta_{13}}), \\
V^*_{4b} & =& - s_{34} - s_{24} e^{-i \delta_{24} } s_{23} 
-s_{14} e^{-i \delta_{14} } s_{13} e^{i \delta_{13}}.
\end{eqnarray}
\end{widetext}

\subsection{ FCNC in {\boldmath $Z^0$} Couplings From Extra Iso-singlet 
Down Quarks}

The FCNC amplitudes are given in terms of the mixings $V_{4i}$ to
form the iso-singlet down quark by \cite{Shin:1989eu}
\begin{equation}
-U_{ij} \equiv V^*_{4i} V_{4j} \quad {\rm for}\quad i \neq j.  
\end{equation}
The FCNC couplings of the down quarks to the $Z^0$ are then given by
\begin{equation}
{\cal L}^Z_{FCNC} = - \frac{e}{2 \sin{\theta_W} \cos{\theta_W}} 
U_{ij} \bar{d}_{iL} \gamma^\mu d_{jL} Z_\mu. 
\end{equation}
The flavor changing neutral currents are
\cite{Nir:1990yq,Silverman:1992fi} $-U_{sd} = V^*_{4s} V_{4d}$ ,
$-U_{sb} = V^*_{4s} V_{4b}$, and $-U_{bd} = V^*_{4b} V_{4d}$.

The diagonal neutral current couplings are reduced in strength by
the amplitudes into the iso-singlet down quarks, becoming
\begin{equation}
{\cal L}^Z_{NC} = -\frac{e}{2\sin{\theta_W}\cos{\theta_W}}
\sum_i (1-|V_{4i}|^2) \bar{d}_{iL} \gamma^\mu d_{iL} Z_\mu.
\label{LNC}
\end{equation}

\begin{figure}
%\centerline{\epsfbox{bwbmix.eps}}
\includegraphics[width=8cm]{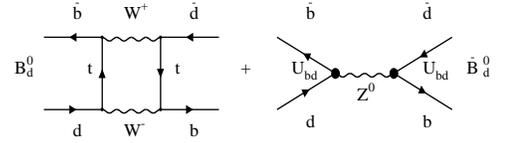}
\caption{\label{bmix} The SM second order weak box diagram plus the 
  double FCNC vertex tree diagram with an intermediate $Z^0$ for
  $B_d-\bar{B_d}$ mixing}
\end{figure}

The FCNC with tree level $Z^0$ mediated exchange may contribute part
of $B_d^0 - \bar{B_d^0}$ mixing and of $B_s^0 - \bar{B_s^0}$ mixing,
and the constraints leave a range of values for the fourth quark's
mixing parameters.  As shown in Fig.\ \ref{bmix}, $B_d^0 - \bar{B_d^0}$
mixing may occur by the $\bar{b} - d$ quarks in a $B_d$
annihilating to a virtual $Z$ through a FCNC with amplitude $U_{bd}\ 
$, and the virtual $Z$ then creating $b - \bar{d}$ quarks through
another FCNC, again with amplitude $U_{bd}$, which then becomes a
$\bar{B_d}$ meson.
  
If the FCNC amplitudes are a large contributor to the $B_d -\bar{B_d}$
mixing, they introduce three new mixing angles and two new phases over
the standard model (SM) into the $CP$ violating $B$ decay asymmetries.
The size of the contribution of the FCNC amplitude $U_{db}$ as one
side of the unitarity quadrangle is less than 0.15 of the unit base
$|V_{cd} V_{cb}|$ at the 1-$\sigma$ level (see Section VI), but we
have found
\cite{Silverman:1996rk,Shin:1989eu,Nir:1990yq,Silverman:1992fi} that
it can contribute as large an amount to $B_d -\bar{B}_d$
mixing as does the standard model.  The new phases can appear in this
mixing and give total phases different from that of the standard model
in $CP$ violating $B$ decay
asymmetries \cite{Nir:1990yq,Silverman:1992fi,Choong:1994rb,
  Branco:1993wr,Lavoura:1993qd}.

For $B_d - \bar{B}_d$ mixing with the four down quark induced $b-d$
coupling, $U_{db}$, we have \cite{Choong:1994rb}
\begin{equation} 
x_d = (2 G_F/3 \sqrt{2}) B_B f_B^2 m_B \eta_B \tau_B \left|
U_{std-db}^2 + U_{db}^2 \right| ,
\end{equation}
where with $y_t = m_t^2/m_W^2$,
\begin{equation}
U^2_{std-db} \equiv  (\alpha/(4 \pi \sin^2{\theta_W})) y_t
f_2(y_t) (V_{td}^* V_{tb})^2,
\label{ustddb}
\end{equation} 
and $x_d = \Delta m_{B_d}/\Gamma_{B_d} = \tau_{B_d} \Delta m_{B_d}$.
In order to compare magnitudes, in the SM, $U^2_{std-db} = 0.50 \times
10^{-6}(1-\rho+i\eta)^2$.

The $CP$ violating decay asymmetries depend on the combined phases of
the $B^0_d-\bar{B}^0_d\;$ mixing and the $b$ quark decay amplitudes
into final states of definite $CP$.  Since we have found that $Z$
mediated FCNC processes may contribute significantly to
$B^0_d-\bar{B}^0_d\;$ mixing, the phases of $U_{db}\ $ would be
important.  
The FCNC amplitude $U_{db}$ to leading order in the
new angles is
\begin{eqnarray}
U_{db} &=& (-s_{34} - s_{24} s_{23} e^{i\delta_{24}}) \nonumber \\
       & & (s_{34} V^*_{td} + s_{14} e^{-i \delta_{14} } -s_{24}
e^{-i \delta_{24} } s_{12}),
\end{eqnarray}
where $V_{td} \approx (s_{12} s_{23} - s_{13} e^{i\delta_{13}})$, and
$V_{ub} = s_{13} e^{-i\delta_{13}}$.

%**********************************************************

\section{Mixing and {\boldmath $CP$} Violating Decay Asymmetries in 
the Four Down Quark Model} 
With new additive contributions to $CP$ violating decay asymmetries,
the asymmetries are no longer sines of SM unitarity triangle angles.
However, they are still sines of the overall phases of the amplitudes
in the asymmetries.  We analyze the FDQM with the present data, and
also show projected results for three different $\sin(2\alpha)$ values
of $-1.0$, 0, and +1.0, which are allowed under the FDQM, although
$\sin(2\alpha) = \pm1$ are not allowed by the SM at 2$\sigma$.

\subsection{{\boldmath $\sin{(2\beta)}$} and {\boldmath $\sin{(2\alpha)}$}}
In the four down quark model we use ``$\sin{(2\alpha)}$'' and
``$\sin{(2\beta)}$'' to denote results of the appropriate $B_d$ decay
$CP$ violating asymmetries, but since the mixing amplitudes are
superpositions, the experimental results for these asymmetries are not
directly related to angles in a triangle. Being imaginary parts of
pure complex exponentials, they are sines of phase angles.  The
asymmetries with FCNC contributions included are (for $\bar{B}$ mixing
to $B$ before decay)
\begin{widetext}
\begin{equation}
\sin{(2\beta)} \equiv A_{B^0_d \to \Psi K^0_s} =
{\rm Im} \left[ \frac{(U^2_{std-db} + U^2_{db})}
{|U^2_{std-db} + U^2_{db}|}
\frac{(V^*_{cb} V_{cs})}{(V_{cb} V^*_{cs})}
\frac{(V^*_{us} V_{ud})}{(V_{us} V^*_{ud})} 
\right]
\label{sinbeta}
\end{equation}
\begin{equation}
\sin{(2\alpha)} \equiv -A_{B^0_d \to \pi^+ \pi^-} =
-{\rm Im} \left[ \frac{(U^2_{std-db} + U^2_{db})}
{|U^2_{std-db} + U^2_{db}|}
\frac{(V^*_{ub} V_{ud})}{(V_{ub} V^*_{ud})} \right]
\label{sinalpha}
\end{equation}
\end{widetext}
with $U^2_{std-db}$ defined in Eq.~(\ref{ustddb}).  The same mixing
phase occurs in both asymmetries, times the squares of the different
decay phases.  We take the Moriond 2002 results for
$\sin{(2\beta)}$ from Babar \cite{Aubert:2002gv} and
Belle \cite{Abe:2001xe,Trabelsi:2002} to give the
weighted average $\sin(2\beta) = 0.78 \pm 0.08$.

%************************* sin(gamma)
\subsection{{\boldmath $\sin{(\gamma)}$}}
In the four down quark model, what we mean by ``$\sin{(\gamma)}$'' is
the result of the experiments which would give this variable in the SM
\cite{Aleksan:1992nh,Aleksan:1991ts}, as in $B^0_s \to D^+_s K^-$.
Here, the four down quark model involves
more complicated amplitudes, and ``$\sin{(\gamma)}$'' is not simply 
$\sin{(\delta_{13})}$:
\begin{equation}
\sin(\gamma) \equiv {\rm Im}   \left[ \frac{(U^2_{std-sb} + U^2_{sb})}
{|U^2_{std-sb} + U^2_{sb}|}
\frac{(V^*_{ub} V_{cs})}{|V^*_{ub} V_{cs}|} 
\frac{(V^*_{cb} V_{us})}{|V^*_{cb} V_{us}|}
\right],
\label{singamma}
\end{equation}
where
\begin{equation}
U^2_{std-sb} \equiv (\alpha/(4 \pi \sin{\theta_W}^2))y_t f_2(y_t)
(V^*_{ts}V_{tb})^2.
\end{equation}
In the SM, $U^2_{std-sb} = 10\times 10^{-6}$.

%****************************  x_s
\subsection{The ``Frequency'' of {\boldmath $B_s$} Oscillations, 
{\boldmath $x_s$}}
In the four down quark model, $x_s$ is no longer the simple ratio of
two CKM matrix elements, but now involves the $Z$-mediated
annihilations and exchange amplitudes as well.  Here we avoid the full
theoretical uncertainty on $B_{B} f_{B}^2$, by taking the ratio of
$x_s$ to $x_d$, which is better calculated theoretically, and in which
we have also included the FCNC with $Z^0$ exchange
\begin{equation}
x_s = \frac{\Delta m_s}{\Gamma_{B_s}} = 1.35 \ x_d 
\frac{ |U^2_{std-sb}+U^2_{sb}| }
{|U^2_{std-db}+U^2_{db}|}.
\end{equation}

We now include the amplitude method analysis of LEP with SLD to assign
a $\Delta \chi^2$ for each $\Delta m_s$ calculated in the angular
parameter grid \cite{lepbosc:2002}.

\subsection{The {\boldmath $B_s$} Decay Asymmetry, 
{\boldmath $\sin{(2\phi_s)}$}}
In the standard model, $B_s$ mixing involves $(V_{ts}^*V_{tb})^2$
which is almost exactly real, and the leading decay process of $b \to
c \bar c s$ has no significant phase from the decay which is
proportional to $V_{cb}^2$.  Thus almost no $CP$ violating phase
develops in the most likely $B_s$ decays.  This occurs in the decays
$B_s \to J/\Psi \phi$, $B_s \to D_s^+ D_s^-$, and $B_s \to J/\Psi
K_S$.  The near vanishing of this asymmetry is a test of the
SM \cite{Nir:1990hj}.  Below, we will find a strange twist on this,
since the FDQM will include a range that includes values smaller than
the SM range, and does not exceed it.  In the SM the angle $\phi_s$ is
the small angle in the $b-s$ unitarity triangle, and its non-zero
value indicates CP violation.

In the four down quark model, the $CP$ violating $B_s$ decay asymmetry is
(for the mixing to $J/\psi~ \phi$ or $D^+_s~ D^-_s$ without the final $K_S$)
\begin{equation}
\sin{(2\phi_s)} = - {\rm Im}\left(\frac{(U_{std-sb}^2 + U_{sb}^2)}
{|U_{std-sb}^2 + U_{sb}^2|} \frac{(V_{cb}^* V_{cs})}{(V_{cb} V_{cs}^*)}
\right),
\label{phis}
\end{equation}
which includes the double FCNC $Z^0$ exchange proportional to
$U^2_{bs}$.  Because of the additional flavor changing term, in the
four down quark model, the angle given by the above asymmetry will not
generally be an angle in a triangle.

%*************************************
\section{Four Down Quark Model Amplitudes in Kaon Experiments}

\subsection{FCNC as an addition to  Penguin plus Box Amplitudes}
Since $CP$ violation and FCNC experiments with $K$ mesons are approaching
the SM range and also limit FCNC amplitudes, we now include the full
SM amplitudes with the FCNC $Z^0$ exchange amplitudes as well.
The $K$ meson experiments are $\epsilon$, $K^+ \to \pi^+ \nu \bar{\nu}$,
$K_L \to \mu \mu$, and we now add the recent and fairly well
determined results for Re$(\epsilon/\epsilon')$.

We use the amplitudes determined by 
Buras \cite{Buras:1999tb,Buras:1998ra}.
In order to reconcile the notation between us and Buras and 
Silvestrini \cite{Buras:1998ed}, we relate their $Z_{ds}$ to our
$U_{sd}$ by taking
\begin{equation}
Z_{ds} = - \left(\frac{\pi^2}{\sqrt{2} G_F m^2_W}\right) U_{sd},
\end{equation}
as implied by their definitions in Lagrangians.

In the following formulas, $B_0$ is the $\Delta S = 1$ box amplitude, 
$S_0$ is the $\Delta S = 2$ box amplitude, $C_0$ is the
$Z^0$ Penguin amplitude, $D_0$ is the off-shell photon penguin (with
$D_0^\prime$ being the on-shell amplitude), and $E_0$ is the 
off-shell gluon penguin ($E_0^\prime$ being on-shell).  Gauge independent
combinations are
\begin{eqnarray}
X_0 & = & C_0 - 4B_0 \\
Y_0 & = & C_0 - B_0  \\
Z_0 & = & C_0 +\frac{1}{4}D_0
\end{eqnarray}
For $m_t = 170$ GeV and $m_c = 1.25$ GeV, for example, these quantities are
$S_0(x_t)= 2.46$, $S_0(x_c)=x_c$, $S_0(x_c,x_t) = 0.0022$,
$X_0=1.57$, $Y_0 = 1.02$,
$Z_0=0.71$, $E_0=0.26$, $D_0^\prime=0.38$, and $E_0^\prime=0.19$.

The FCNC  $Z^0$ exchange with amplitude $U_{ds}$ can be added to the 
$d-s$ Penguin amplitude with the $Z^0$ by the substitution
\begin{equation}
\lambda _{t}C_{0}\left(x_{t}\right) 
\rightarrow 
\lambda _{t}C_{0}\left( x_{t}\right) -\frac{\pi ^{2}}
{\sqrt{2}G_{F}M_{W}^{2}}U_{sd},
\label{subs}
\end{equation}
to obtain the SM plus FCNC result.

\subsection{Indirect {\boldmath $CP$} Violation in Epsilon}
In $K-\bar{K}$ mixing, the small indirect $CP$ violation
is given through $|\epsilon|$ \cite{Buras:1999tb} 
%(From Buras, hep-ph/9905437, page 19 equation 3.22 
%and page 21 equation 3.36.)
\begin{equation}
\left| \epsilon \right| =\frac{1}{\sqrt{2}\Delta M_{K}}|{\rm Im}M_{12}|,
\end{equation}
where we include the substitution from Eq.~\ref{subs}
\begin{widetext}
\begin{eqnarray}
M_{12} &=& \frac{G_{F}^{2}}{12\pi ^{2}}F_{K}^{2}\widehat{B}%
_{K}m_{K^{0}}M_{W}^{2}\left[ \lambda _{c}^{*2}\eta _{1}S_{0}(x_{c}) 
+\lambda _{t}^{*2}\eta _{2}S_{0}(x_{t}) 
+2\lambda_{c}^{*}\lambda _{t}^{*}\eta_{3}S_{0}(x_{c},x_{t}) \right] 
\nonumber \\
 & & -\frac{\sqrt{2} G_F}{12} F_K^2 \widehat{B}_K m_{K^0} U_{sd}^2.
\end{eqnarray}
\end{widetext}
The short distance QCD corrections factors in NLO are \cite{Buras:1999tb}
$\eta_1 = 1.38 \pm 0.20$, $\eta_2 = 0.57 \pm 0.01$, and 
$\eta_3 = 0.47 \pm 0.04$, and we use $\hat{B_K} = 0.85 \pm 0.13$.

\subsection{Direct {\boldmath $CP$} Violation in 
  Re{\boldmath $(\epsilon'/\epsilon)$}} 
The direct $CP$ violation in
$K^0$ decays, Re$(\epsilon'/\epsilon)$, has received more accurate
measurements that are definitely non-zero.  The average of KTeV
\cite{Yamanaka:2001tq} and NA48 \cite{Vallage:2001vc} gives
Re$(\epsilon'/\epsilon) = 17.3 \pm 2.4$, where the error has been
increased by $\sqrt{(\chi^2/df)}$.  The sum of the SM
\cite{Buras:1999tb,Buras:1998ra} plus FCNC amplitude from
Eq.~\ref{subs} is
%(From Buras, hep-ph/9905437, page 30 equation 5.5.)
\begin{equation}
{\rm Re}(\epsilon'/\epsilon)=
{\rm Im}\lambda _{t}F_{\epsilon^{^{\prime }}} 
- \frac{\pi ^{2}}{\sqrt{2}G_{F}M_{W}^{2}}
{\rm Im}U_{sd}\left[
P_{X}+P_{Y}+P_{Z}\right],
\end{equation}
where
\begin{equation}
F_{\epsilon ^{^{\prime }}}=P_{0}+P_{X}X_{0}(x_{t})
+P_{Y}Y_{0}(x_{t}) +P_{Z}Z_{0}(x_{t})
+P_{E}E_{0}(x_{t}).
\end{equation}
The P's are functions of $B^{1/2}_6 = 1.0\pm 0.3$, $B^{3/2}_8 = 0.8
\pm 0.2$, and $\Lambda^{(4)}_{\overline{MS}} = 340 \pm 50$ MeV
\cite{Buras:1999tb}.

\subsection{{\boldmath $K^+ \to \pi^+ \nu \overline{\nu}$}}
The recent detection of two events in $K^+ \to \pi^+ \nu \bar{\nu}$
has produced the experimental 
result \cite{Adler:2001xv}
\begin{equation}
{\rm BR_{expt}}(K^+ \to \pi^+ \nu \bar{\nu}) = 
(1.57^{+1.75}_{-0.82})\times 10^{-10}
\end{equation}
compared to the SM range \cite{D'Ambrosio:2001zh} of $(0.72 \pm
.21)\times 10^{-10}$.  The Poisson probability for the angle
parameters is converted to a chi-squared form \cite{Silverman:1998tw}
which is convolved into the total $\chi^2$ formula.  For this
experiment using a logarithmic prior, with $2\times n_{obs}=4$ degrees
of freedom \cite{Silverman:1998tw}, the addition to $\chi^2$ is
\begin{equation}
\chi^2 = 2 <n> = 2 \times ({\rm 2~events}) \times 
\frac{\rm BR_{calc}}{\rm BR_{expt}}.
\end{equation}
The sum of the SM \cite{Buras:1999tb}
plus FCNC contributions is obtained by using Eq.~(\ref{subs})
%(From Buras, hep-ph/9905437, page 43, equations 6.7-6.9.)
%\begin{widetext}
\begin{eqnarray}
BR_{\rm calc} & (K^+ & \rightarrow  \pi^+ \nu \overline{\nu }) 
= r_K BR( K^+ \rightarrow \pi^0 e^+ \nu )  \\
 & \times & \frac{\alpha^2}{|V_{us}|^2 2\pi^2 \sin^{4}{\theta_{W}} }  
\nonumber \\
 & \times & \left[   
2\left|
\lambda_{c} X_{NL}^{e} + \lambda_{t} X(x_{t}) 
-\frac{\pi^2}{\sqrt{2}G_{F}M_{W}^{2}} U_{sd}
\right|^{2} \right. \nonumber \\ 
  &+ &  
\left. \left| \lambda_{c} X_{NL}^{\tau} + \lambda_{t} X(x_{t}) 
-\frac{\pi^{2}}{\sqrt{2}G_{F}M_{W}^{2}} U_{sd} 
\right|^{2}
\right] \nonumber.
\end{eqnarray}
%\end{widetext}
Here, $X(x_t) = \eta_x X_0(x_t)$, etc., and \cite{Buchalla:1998ba} 
$\eta_x = 0.994$.
Without the SM, the contribution of the  $Z^0$
exchange alone with amplitude $U_{sd}$ is 
$\chi^2 = 1.61 \times 10^9 |U_{sd}|^2$.

\subsection{{\boldmath $K_L \to \mu^+ \mu^-$}} 
The short distance weak FCNC contribution to $K_L \to \mu \mu$
constrains Re$(U_{ds})$ and is given from \cite{Buras:1997fb}
after including Eq.~\ref{subs}
%(From Buras hep-ph/9704376, page 102 equations 7.71 and 7.72.)
%\begin{widetext}
\begin{eqnarray}
& BR &(K_{L}\rightarrow \mu ^{+}\mu ^{-})
  = 
BR(K^{+}\rightarrow \mu^{+}\nu) \nonumber \\ 
 & \times & \frac{\tau_{K_{L}}}{\tau_{K^{+}}}\frac{\alpha ^{2}}
{|V_{us}| ^{2}\pi ^{2}\sin ^{4}\theta _{W}}  \\
 & \times & 
\left( 
{\rm Re}\lambda_{c}Y_{NL}+{\rm Re}\lambda_{t}Y(x_{t})
 - \frac{\pi^{2}}{\sqrt{2}G_{F}M_{W}^{2}}{\rm Re}U_{sd}
\right)^{2}. \nonumber
\end{eqnarray}
%\end{widetext}
Here, $Y(x_t)=\eta_y Y_0$, and \cite{Buchalla:1998ba} $\eta_Y = 1.012$.
The long distance contribution has been
analyzed \cite{D'Ambrosio:1998jp}.  We make the 1$\sigma$ limit
conservatively as the sum of the 1$\sigma$ experimental limit plus the
1$\sigma$ long distance estimate \cite{D'Ambrosio:1998jp}.

From the above $K$ meson formulas, the error formulas were generated
using Mathematica.

%*************************************
\section{Joint Chi-squared Analysis of the SM and the FDQM Experiments}

FCNC experiments put limits on the new mixing angles and constrain the
possibility of new physics contributing to $B_d^0 - \bar{B_d^0}$ and
$B_s^0 - \bar{B_s^0}$ mixing.  Here we jointly analyze all constraints
on the $4 \times 4$ mixing matrix obtained by assuming only one of the
SU$(2)_L$ singlet down quarks mixes appreciably \cite{Nir:1990yq}.  We
use the seven experiments for the $3 \times 3$ CKM sub-matrix elements
\cite{Choong:1994gq}, which include: those on the three matrix
elements $V_{us}, V_{ub}, V_{cb}$ of the $u$ and $c$ quark rows;
$|\epsilon|$; $B_d-\bar{B}_d$ mixing ($x_d$); the new limits on
$\Delta m_s$, or $x_s$; and the new measurements for $\sin{(2\beta)}$.
For studying FCNC, we include $V_{ud}$ and $V_{cd}$, the bound on $B
\to \mu \mu X_s$ (which constrains $b \to s$), the two events in $K^+
\to \pi^+ \nu \bar{\nu} $ \cite{Silverman:1998tw,Lavoura:1993qd,
  Adler:2000by}, and $R_b$ in $Z^0 \to b \bar{b}$
\cite{Lavoura:1993qd} (which directly constrains the $V_{4b}$ mixing
element).  FCNC experiments will bound the three amplitudes $U_{ds}$,
$U_{sb}$, and $U_{bd}$ which contain three new mixing angles and three
phases.  We use the mass of the top quark as $m_t = 174$ GeV.  We also
add FCNC constraints from $K_L \to \mu \mu$, now including the large
long distance error, and the new and more convergent results for
$\epsilon'/\epsilon$ from NA48 \cite{Vallage:2001vc} and
KTeV \cite{Yamanaka:2001tq}.

Related analyses including both SM and FDQM amplitudes in kaon
constraints by Barenboim, Botella and
Vives \cite{Barenboim:2000zz,Barenboim:2001fd} preceed this work.  We
have applied a full $\chi^2$ analysis rather than just 95\% CL bounds,
and have included the new, larger and more exact $\sin{(2\beta)}$
results, as well as new $K^+ \to \pi^+ \nu \bar{\nu}$,
$\epsilon'/\epsilon$ results, and new and full $x_s$ data.  We have
also included an analysis of the $4 \times 4$ mixing matrix parameters
and found a crucial cancellation in one of the matrix elements.
 
We use a method for combining the Bayesian Poisson distribution for
the average for the two observed events in $K^+ \to \pi^+ \nu
\bar{\nu}$ \cite{Adler:2001xv, Silverman:1998tw} with the chi-squared
distribution from the other experiments.  This treats the two events
with a logarithmic Bayesian prior as four degrees of freedom.  This
gives a total of ten additional experimental degrees of freedom for
the FDQM.

In maximum likelihood correlation plots, we use for axes two output
quantities which are dependent on the mixing matrix angles and phases,
such as $(\rho,\eta)$, and for each possible bin with given values for
these, we search through the nine dimensional angular data set of the
$4 \times 4$ down quark mixing angles and phases, finding all sets
which give results in the bin, and then put into that bin the minimum
$\chi^2$ among them.  To present the results, we then draw contours at
several $\chi^2$ in this two dimensional plot corresponding to given
confidence levels.

%********************* SM sin2alpha-sin2beta plots
\subsection{Standard Model {\boldmath $(\sin{(2\alpha)},\sin{(2\beta)})$} 
  Plot - Present Constraints} 
\begin{figure}%[ht]
%\epsfxsize = 6in
%\centerline{\epsfbox{smckmab.eps}}
\includegraphics[width=8cm]{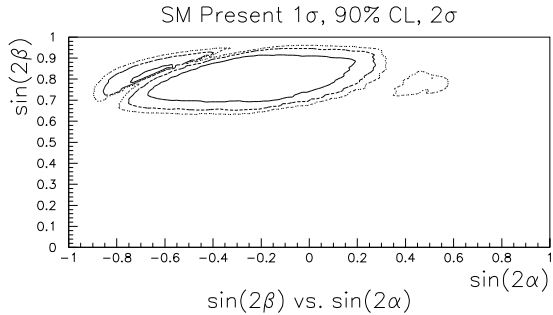}%
\caption{\label{sa-sbSM}
The ($\sin{(2\alpha)}$, $\sin{(2\beta)}$) plot for the standard model
with contours at 1$\sigma$, 90\%CL, and 2$\sigma$ with present data.}
\end{figure}

For the SM we take $\lambda = V_{us}$ as
fixed and then use the six experiments on the $3\times3$ CKM matrix
elements named above with three parameters to give three degrees of
freedom.  In the figures we show the $\chi^2$ contours with confidence
levels (CL) at values equivalent to 1$\sigma$, 90\% CL (1.64$\sigma$),
and 2$\sigma$.  The new BaBar \cite{Aubert:2001nu} and Belle
\cite{Abe:2001xe,Trabelsi:2002} average is $\sin{(2\beta)}= 0.78 \pm
0.08$.  This gives $\beta = 25.6^{\circ + 4.0^\circ}_{ - 7.4^\circ}$.
From Fig.\ \ref{sa-sbSM} for the SM we see that the $\sin(2\beta)$
range is from 0.63 to 0.96 at 2$\sigma$, centered around the
experimental average of 0.78.  The SM $\sin(2\alpha)$ range at
2$\sigma$ is from -0.90 to +0.57.

%************************ FDQM sin(2alpha) sin(2beta) plots
\subsection{Four Down Quark Model {\boldmath $(\sin{(2\alpha)},
    \sin{(2\beta)})$} Plots - Present Limits} 
\begin{figure}%[ht]
%\epsfxsize = 6in
%\centerline{\epsfbox{proj4prlog.eps}}
\includegraphics[width=8cm]{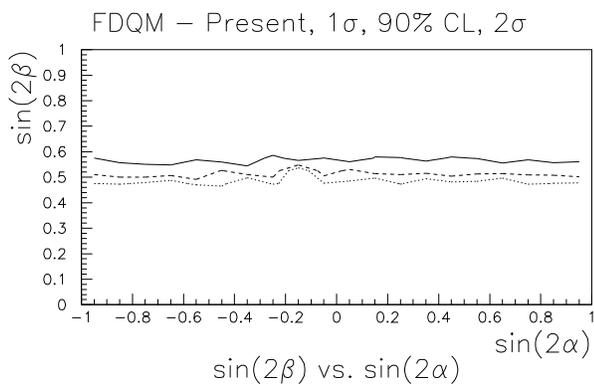}%
\caption{\label{sa-sb4q} The $(\sin(2\alpha),\sin(2\beta))$ plot for the
  FDQM with contours at 1$\sigma$, 90\% CL, and 2$\sigma$.}
\end{figure}

In the FDQM analysis
including FCNC experiments, there are 17 experimental degrees of
freedom, minus 9 parameters, giving 8 remaining degrees of freedom.
In contrast to the SM, for the FDQM in Fig.\ \ref{sa-sb4q}, almost the
entire region $\sin(2\beta) > 0.48$ is allowed at 2$\sigma$, and
$\sin(2\beta)$ can be as low as 0.55 at 1$\sigma$.  In the FDQM, all
values of $\sin(2\alpha)$ are allowed.  In this case, the larger
1$\sigma$ range for $\sin{(2\beta)}$ than from the direct experimental
measurement is an effect of including so many experiments in the joint
fit.

%******************** SM comparison rho-eta plots
\subsection{Standard Model with Comparison Experiments 
  {\boldmath ($\rho,\eta$)} Plot} 
\begin{figure}%[ht]
%\epsfxsize = 6in
%\centerline{\epsfbox{bwckmre.eps}}
\includegraphics[width=8cm]{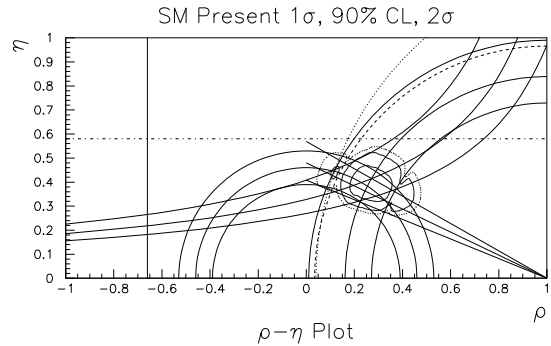}%
\caption{\label{recompSM}  The $(\rho,\eta)$ plot for the SM with three
comparison kaon experiments added, with joint $\chi^2$ contours at
1$\sigma$, 90\%CL, and 2$\sigma$.  The light lines are for the Kaon
experiments and are described in the text.} 
\end{figure}

Here we depart from the analysis of
the SM experiments alone to show the effects of the additional three
$K$ meson experiments, namely $K^+ \to \pi^+ \nu \bar{\nu}$, 
$K_L \to \mu \mu$, and $\epsilon'/\epsilon$.
While they are not needed in the SM analysis,
they are included in the FDQM analysis.  In this case, there are
13 experimental degrees of freedom, minus 4 parameters, 
giving a net 9 degrees of freedom.
Contours are at 1$\sigma$, 90\% CL, and 2$\sigma$.  The three new $K$
meson experimental contours for the SM are shown in
Fig.\ \ref{recompSM}.  For $\epsilon'/\epsilon$, the lower 1$\sigma$
contour is the horizontal dot-dashed line, where the central contour
would be a horizontal line at $\eta = 1.1$.  For $K_L \to \mu \mu$ the
solid vertical line at $\rho = - 0.66$ is the lower 1$\sigma$ contour
with the central contour being a vertical line at $\rho = 0.9$, which
is not shown.  This includes conservatively a large and uncertain long
distance effect \cite{D'Ambrosio:1998jp}.  For $K^+ \to \pi^+ \nu
\bar{\nu}$ the dotted contour giving an additional $\chi^2 = 1$ is the
arc of the circle centered about $(\rho,\eta) = (1.3,0)$.  It is not
quite as restrictive in the SM as the 90\% CL from the $x_s$ or
$\Delta m_s$ limit, which is shown as the dotted quarter-circle about
$(\rho, \eta) = (1.0, 0)$.  For the rest of the SM analysis we drop
these three new $K$ meson experiments.

%*************************  SM rho-eta plots
\subsection{Standard Model: {\boldmath ($\rho,\eta$)} Plot}
\begin{figure}%[ht]
%\epsfxsize = 6in
%\centerline{\epsfbox{bwsmckmre.eps}}
\includegraphics[width=8cm]{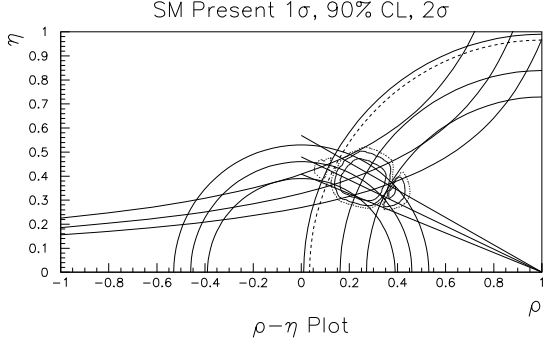}%
\caption{\label{rho-etaSM} 
  The $(\rho,\eta)$ plot for the standard model, showing the
  1$\sigma$, 90\% CL, and 2$\sigma$ contours of the joint fit, and the central
and 1$\sigma$ contours of the various constraints.}
\end{figure}

For the SM $(\rho,\eta)$ plot in Fig.\ \ref{rho-etaSM}, the joint
$\chi^2$ enclosed contours are at 1$\sigma$, 90\% CL, and 2$\sigma$.
The half-circles about $(\rho, \eta)=(0,0)$ are the center and
1$\sigma$ contours for $|V^*_{ub}V_{ud}/V^*_{cb}V_{cd}|$.  The
hyperbolas are the center and 1$\sigma$ contours for $\epsilon$.  The
quarter circles about $(\rho,\eta) = (1,0)$ are for $|V_{td}|$ from
$x_d$ in $B_d - \bar{B_d}$ mixing.  The lines emanating from
$(\rho,\eta)=(1,0)$ are the central and 1$\sigma$ limits for
$\sin{(2\beta)}$.
The $\Delta m_s$ 90\% circular arc is shown as the dashed
quarter-circle, although the analysis weights each $\Delta m_s$ or
each $V_{td}$ in $\chi^2$.  We see the effects of the $x_s = (\Delta
m_s/\Gamma_s)= 1.35 x_d |V_{ts}/V_{td}|^2$ lower bound in the SM
limiting the length of $V_{td} \propto \sqrt{(1-\rho)^2+\eta^2}$ and
cutting off $\rho$ for $\rho < 0$.

%******************************  FDQM rho-eta plots
\subsection{Four Down Quark Model: {\boldmath ($\rho,\eta$)} Plots}
As in the SM, the plotted $\rho$ and $\eta$ are taken as the 
coordinates of $V_{ub}^*$, scaling the base of the $b-d$ 
unitarity quadrangle to unity 
\begin{equation}
\rho + i \eta \equiv V^*_{ub}V_{ud}/|V^*_{cb} V_{cd}|.
\end{equation}
The unitaritity quadrangle is given by
\begin{equation}
V^*_{ub}V_{ud}+V^*_{cb}V_{cd}+V^*_{tb}V_{td}+V^*_{4b}V_{4d}=0
\end{equation}
where the last term has limits $|U_{bd}/V^*_{cb} V_{cd}| \leq 0.15$,
as will be shown later.  The near half circles in $\gamma = \delta =
\delta_{13}$ ($V^*_{ub} = s_{13} e^{i\delta_{13}}$) at present are due
to $\delta_{14}$ or $\delta_{24}$ (which are related) becoming some of
the source of the observed $CP$ violation in $\epsilon$, so that
$\delta_{13}$ is less constrained.  Then, $\delta_{13}$ can be closer
to zero or $180^\circ$ so that $\eta$ can also be small or zero.  For
projected $\sin(2\alpha)=+1$, 0, or $-1$, we see regions extended
beyond the SM regions, which also allow $\eta$ to be small.  Examining
the effect of each new $K$ experiment separately, we find that the
$K^+ \to \pi^+ \nu \bar{\nu}$ result eliminates the large 1$\sigma$
negative $\eta$ rings from the previous
analysis \cite{Silverman:1998uj}.
\begin{figure}%[ht]
%\epsfysize = 5in
%\centerline{\epsfbox{fullfqcomblogre.eps}}
\includegraphics[width=8cm]{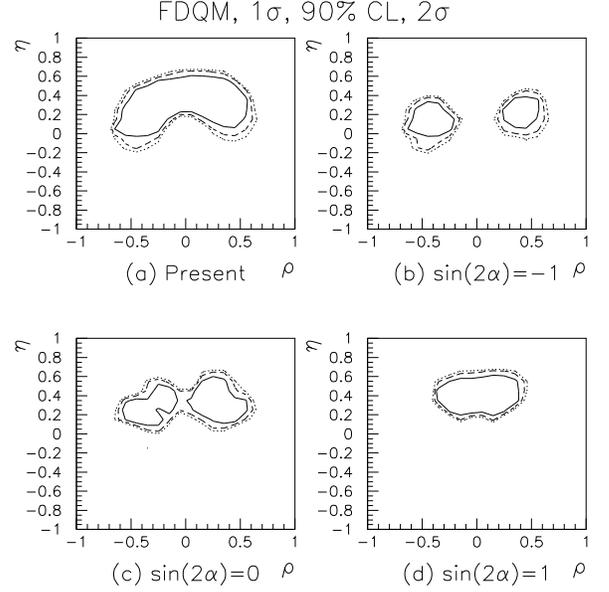}%
\caption{\label{rho-eta4q}
The $(\rho,\eta)$ plots for the four down quark model from:
(a) present data, and for projected
$\sin{(2\alpha)}$ values of $-1$, $0$, and $1$.  
Contours are at 1$\sigma$, 90\% CL, and 2$\sigma$.}
\end{figure}

%***************** epsilon-FCNC
\subsection{Fraction of the New FCNC Amplitude in {\boldmath $\epsilon$}}
\begin{figure}%[ht]
%\epsfxsize = 6in
%\centerline{\epsfbox{proj6prlog.eps}}
\includegraphics[width=8cm]{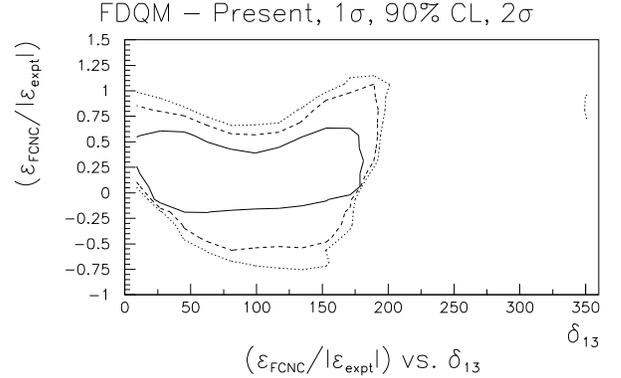}%
\caption{\label{repsilon}
  The ratio $(\epsilon_{\rm FCNC}/|\epsilon_{\rm expt}|)$ of the
  contribution of the FCNC amplitude to $\epsilon_K$ as a function of
  the angle $\delta_{13}$.}
\end{figure}

In order to display how the FCNC $Z^0$ exchange with the new phases in
$U_{ds}$ can account for the $CP$ violation in $\epsilon_K$, we plot
the ratio of the FCNC contribution to the experimental result.  In
Fig.\ \ref{repsilon} $(\epsilon_{\rm FCNC}/|\epsilon_{\rm expt}|)$ is
shown against the phase of $V^*_{ub}$, which is $\delta_{13}$.  In
Fig.\ \ref{repsilon}, while $\epsilon_{\rm FCNC}$ cannot account for
the entire $\epsilon$ result, it can account for 60\% of it at a
1$\sigma$ confidence level.

%****************** SM xs sin(gamma) plot
\subsection{Standard Model: {\boldmath $(x_s,\sin{(\gamma)})$} Plots}
$x_s$ is
determined in the SM from 
\begin{equation}
x_s = 1.35 x_d (|V_{ts}|/|V_{td}|)^2.  
\end{equation}
The
largest error arises from the uncertainty in $|V_{td}|$, which follows
from the present $15\%$ uncertainty in $\sqrt{B_B} f_B = 230 \pm 35$
MeV from lattice calculations \cite{Bernard:2000ki}.  
In the SM, the $B$ factory measurements
construct a rigid triangle from the knowledge of $\alpha$ and
$\beta$, and removes this uncertainty in $\gamma$ and $x_s$ in the
future.  

\begin{figure}%[ht]
%\epsfysize = 5in
%\centerline{\epsfbox{smckmxsg.eps}}
\includegraphics[width=8cm]{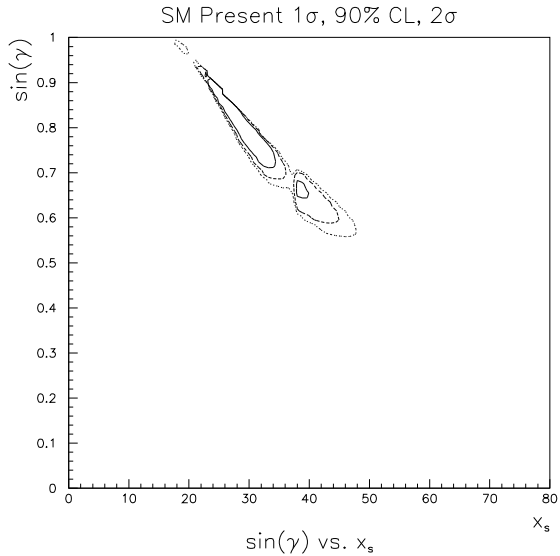}%
\caption{\label{xs-sgSM}
The ($x_s$, $\sin{\gamma}$) plot for the standard model
with present limits with contours at 1$\sigma$, 90\%CL, and 2$\sigma$.} 
\end{figure}

From present data for the SM $(x_s,\sin{(\gamma)})$ plot 
in Fig.\ \ref{xs-sgSM}, the limits at 2$\sigma$ are 
$0.56 \le \sin{(\gamma)} \le 0.99$, and $16 \le  x_s \le 48$.
Because of the approximately linear relation between $x_s$ and 
$\sin(\gamma)$, an exact $x_s$ measurement ($\propto1/|V_{td}|^2$) 
can strongly constrain $\sin{(\gamma)}$ to $\pm 0.07$ in the SM.

%*************************** FDQM xs sin(gamma) plot
\subsection{Four Down Quark Model: {\boldmath 
$(x_s,\sin{(\gamma)})$} Plots}
\begin{figure}%[ht]
%\epsfysize = 5in
%\centerline{\epsfbox{fullfqcomblogxsg.eps}}
\includegraphics[width=8cm]{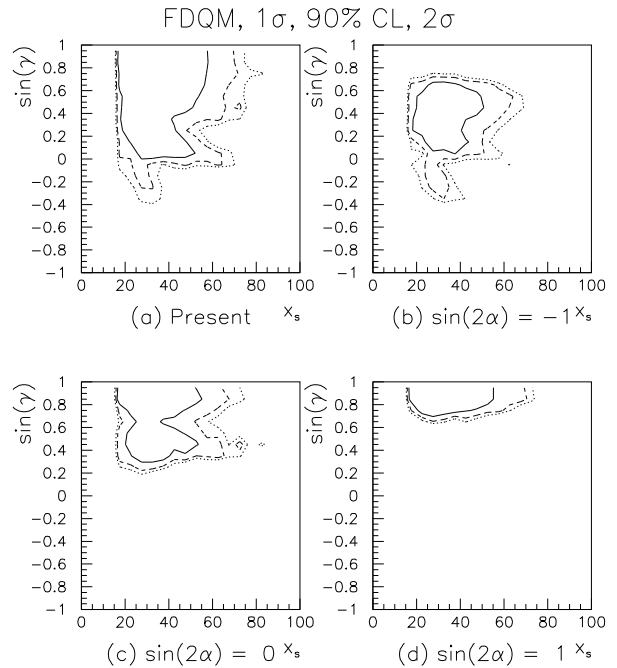}%
\caption{\label{xs-sg4q}
The $(x_s,\sin{(\gamma)})$ plots for the four down quark
model from (a) present data, and (b, c, and d) for $B$ factory cases
for values of $\sin{(2\alpha)}$ as labeled.}
\end{figure}

In the FDQM, the $\sin{(\gamma)}$ range goes down to zero at
1$\sigma$, or $-0.4$ at 2$\sigma$ (Fig.\ \ref{xs-sg4q}), since $\eta$
now goes down to zero at 1$\sigma$ or to $-0.2$ at 2$\sigma$ where
$\rho \approx -0.5$.  A larger $\sin{(\gamma)}$ range is thus allowed
in the FDQM than in the SM.  The $x_s$ allowed region in the FDQM is
16 to 60 at 1$\sigma$ or to 80 at 2$\sigma$, which is also larger than
the 2$\sigma$ $x_s$ range of 48 in the standard model.  In the FDQM,
there is not an approximately linear relation between $\sin(\gamma)$
and $x_s$ as there is in the SM.  Thus an accurate measurement of
$x_s$ still leaves a very large region of $\sin(\gamma)$ available in
the FDQM.  A subsequent $\sin(\gamma)$ measurement will be needed to
distinguish between the two models.

%********************* sin(2\phi_s) plot
\subsection{The Decay Asymmetry from {\boldmath $B_s$} 
Mixing, {\boldmath $\sin{(2\phi_s)}$}}
\begin{figure}%[ht]
%\epsfysize = 4in
%\centerline{\epsfbox{proj8prlog.eps}}
\includegraphics[width=8cm]{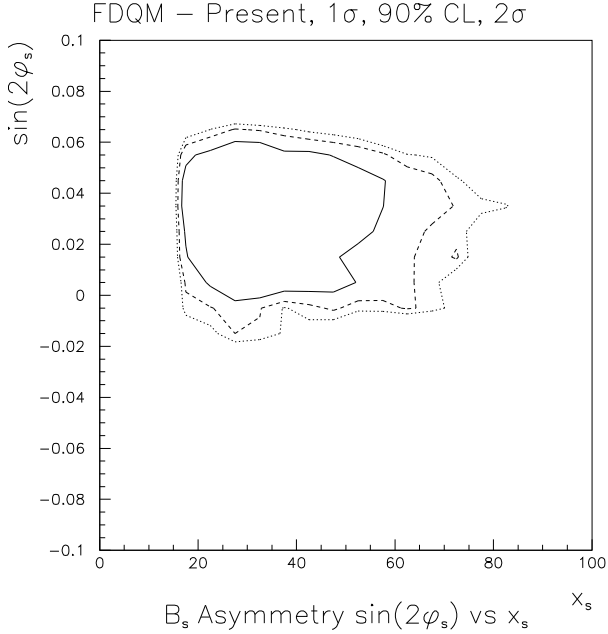}%
\caption{\label{xs-Bs4q}
The $(x_s,\sin{(2\phi_s)})$ plot for the $B_s$ asymmetry 
$\sin{(2\phi_s)}$ in
the four down quark model for present data, with contours at 1$\sigma$, 
90\%CL, and 2$\sigma$.}
\end{figure}

$\phi_s$ is the small angle in the $b-s$ unitarity triangle given
by
\begin{equation}
V^*_{cb}V_{cs}+V^*_{tb}V_{ts}+V^*_{ub}V_{us}=0.
\end{equation}
In Wolfenstein terms this is 
\begin{equation}
(A\lambda^2)\times 1 
+ 1\times (-A\lambda^2 - A\lambda^4(\rho+i\eta)) + 
(A\lambda^3(\rho+i\eta))\times \lambda = 0.
\end{equation}
Then, $\sin{(\phi_s)} = A\lambda^4 \eta/A\lambda^2 = \lambda^2 \eta$.
This is small in the Standard Model where 
$\sin{(2\phi_s)} = 2 \lambda^2 \eta = 0.10 \eta$, or at 2$\sigma$
\begin{equation}
0.030 \leq \sin{(2\phi_s)} \leq 0.060.
\end{equation}
In the FDQM, as seen in Fig.\ \ref{xs-Bs4q}, $-0.2 \leq \sin{(2\phi_s)}
\leq 0.065$ at 2$\sigma$, and down to zero at 1$\sigma$.  Here the
range continues down to zero since $\eta$ can go down to zero.  Hence,
a value of $\sin{(2\phi_s})$ less than 0.03 would signify a deviation
from the SM.

%********************** U_bd plot
\subsection{Fourth Side of the Unitarity Quadrangle {\boldmath $U_{bd}$}}
\begin{figure}%[ht]
%\epsfxsize = 5in
%\centerline{\epsfbox{proj3prlog.eps}}
\includegraphics[width=8cm]{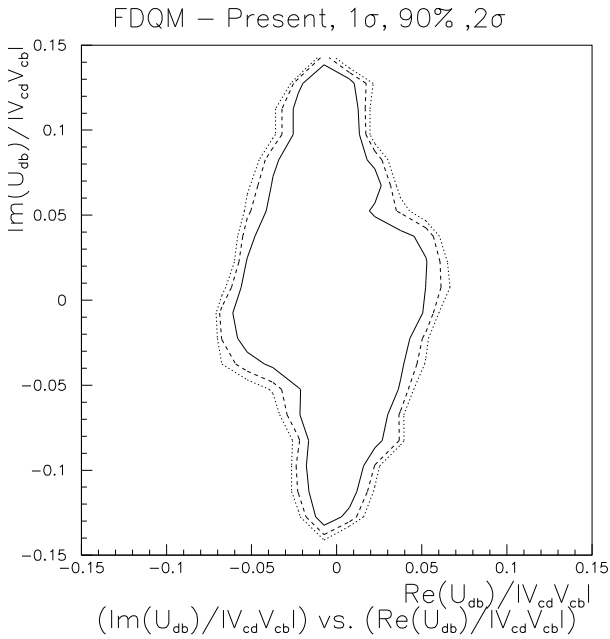}%
\caption{\label{Udb} The complex plot of $U_{db}$ scaled to make it the
fourth side of the unitarity quadrangle in the $b-d$ unitarity plot.}
\end{figure}

Unitarity of the $b-d$ columns has four terms, which may be written
as 
\begin{equation}
V^*_{ub}V_{ud} + V^*_{cb}V_{cd} + V^*_{tb}V_{td} - U_{bd} = 0,
\end{equation}
since $-U_{bd} = V^*_{4b}V_{4d}$.  (We use $U_{db} = U^*_{bd}$).  As
the unitarity triangle is scaled by $|V^*_{cb}V_{cd}|$ to make a unit
base, the complex plot of $U_{bd}$ is also so scaled.  The length of
the $U_{bd}/|V^*_{cb}V_{cd}|$ side, as plotted in Fig.\ \ref{Udb}, is
thus less than 0.15, compared to the unit base in the $(\rho,\eta)$
plot, and prefers possibly a more vertical direction.  The accuracy of
angles and sides of the unitarity triangle must and should reach this
accuracy for a good test of the SM.

%******************  Mixing Matrix Elements
\section{Size of the Mixing Matrix Elements and Mixing Angles}
\subsection{Bound From {\boldmath $Z^0 \to b \bar{b}$} }
The weak isovector part of $Z^0 \to b \bar{b}$ is reduced by 
$(1 - |V_{4b}|^2)$ through
\begin{eqnarray}
V^b &=& -\frac{1}{2}(1 - |V_{4b}|^2)
+\frac{2}{3}\sin^2{\theta_W}+\frac{1}{3}\rho_t, \nonumber \\
A^b &=&-\frac{1}{2}(1-|V_{4b}|^2)
+\frac{1}{3}\rho_t, \\
\Gamma^{std}(Z^0 \to b \bar{b}) &=& \frac{C_{QCD}G_F m_Z^3}{6
\sqrt{2} \pi} \left[(V^b)^2+(A^b)^2 \right],
\end{eqnarray}
where $C_{QCD} = 3(1.0385)$, and
$\rho_t = 0.0094$ for $m_t = 174$ GeV.  
Present data and theory give
\begin{eqnarray}
 R_b^{\rm exp}    &=& 0.21642 \pm 0.00073 \\
 R_b^{\rm theory} &=& 0.2158 \pm 0.0003  
\end{eqnarray}
We note that the $|V_{4b}|$ effect is to decrease $R_b$, while the
experiment is about 1$\sigma$ above the theory.  
To lowest order in $|V_{4b}|$ the FDQM effect is
\begin{equation}
\frac{\Gamma_{b \bar{b}}^{std+FCNC}}{\Gamma_{b \bar{b}}^{std}} =
(1-2.29|V_{4b}|^2).
\end{equation}
This gives a contribution to $\chi^2$ of
\begin{equation}
\chi^2 = (0.82 + 0.68 \times 10^3 |V_{4b}|^2)^2.
\end{equation}
This $\chi^2$ is used as a constraint on all angle choices in the fit.
Taking the 90\% CL limit at $\chi^2 = (1.64)^2$, gives the bound
on $|V_{4b}|$ from $R_b$ alone of $|V_{4b}| \approx |s_{34}| \leq 0.035$.

\subsection{3-D Matrix Element Lego Plot}
\begin{figure}%[ht]
%\epsfysize = 5in 
%\centerline{\epsfbox{projhprlog.eps}}
\includegraphics[width=8cm]{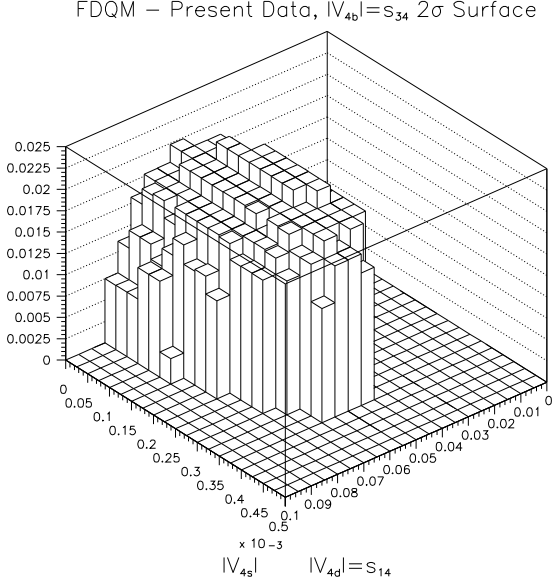}%
\caption{\label{3D} The Lego plot for the height $|V_{4b}|\approx s_{34}$ 
  at 2$\sigma$, on the base of $|V_{4s}|$ (units $10^{-3}$) vs. 
$|V_{4d}|\approx s_{14}$.}
\end{figure}

From the 2$\sigma$ surface in the 3D space of the magnitudes of the
matrix elements involved in the FCNC, Fig.\ \ref{3D}, we can see the
limits and ranges of two of the matrix elements.  To 5\% accuracy,
$V_{4d}=-s_{14}e^{i\delta_{14}}$, and its magnitude ranges from 0.035
to 0.085 at 2$\sigma$.  To 10\% accuracy, $V_{4b}=-s_{34}$, and its
magnitude ranges up to 0.020 at 2$\sigma$.  The third FCNC matrix
element, $|V_{4s}|$, is bounded by 0.0004.  This requires a fine
cancellation between its two components in $( -s_{24}e^{i\delta_{24}}
- s_{12} s_{14}e^{i\delta_{14}})$, such that $s_{24}\approx
s_{12}s_{14}$ and $\delta_{24}=\delta_{14}+\pi$ to get the cancelling
minus sign.  This means that there is effectively only one new phase,
which we may consider as $\delta_{14}$.  From the cancellation,
$s_{24}$ ranges from 0.009 to 0.017 at 2$\sigma$.  The cancellation is
to about $1/20$ of the value of $s_{24}$.  The third term in $V_{4s}$,
$s_{34} (s_{23}+s_{12}s_{13}e^{i\delta_{13}})$, then contributes $\le
0.0009$, which is the same order as the partly canceling terms.  The
cancellation does not mean fine tuning since one could have
parametrized $V_{4s}$ by a single angle instead.  However, the
incredibly small size of $|V_{4s}|\leq \lambda^5$ could be considered
a fine tuning itself. In comparison to the SM CKM matrix we should
note that keeping the leading terms in the real and imaginary parts,
$V_{cs} = 1 +iA^2 \lambda^6 \eta$, $V_{cd} = - \lambda
-iA^2\lambda^5\eta$, and $V_{ts} = -A \lambda^2 -iA\lambda^4\eta$.  So
even in the standard model there are matrix elements whose imaginary
parts are as small as O($\lambda^4$), O($\lambda^5$), and
O($\lambda^6$).

In Wolfenstein terms, $|V_{4d}|\approx s_{14}\approx \lambda^2$,
$s_{24}\approx s_{12} s_{14} \approx \lambda^3$, 
$|V_{4b}|\approx s_{34} <\lambda^2/2$, but
$|V_{4s}|\leq \lambda^5$.  The sequence may violate the heirarchical
expectation from the $3\times3$ CKM matrix.

In the double FCNC $Z^0$ exchange amplitude in $B_d-\bar{B_d}$ mixing,
via
\begin{equation}
U_{db} =- V^*_{4d}V_{4b} \approx  s_{14}e^{-i\delta_{14}} s_{34},
\end{equation}
it is only the $\delta_{14}$ phase in $(U_{db})^2\approx
e^{-2i\delta_{14}}$ that can add to the SM box diagram term with its
phase of $(V^*_{td})^2$.

\subsection{Phases}
\begin{figure}%[ht]
  %\epsfxsize = 4in 
%\centerline{\epsfbox{proj7prlog.eps}}
\includegraphics[width=8cm]{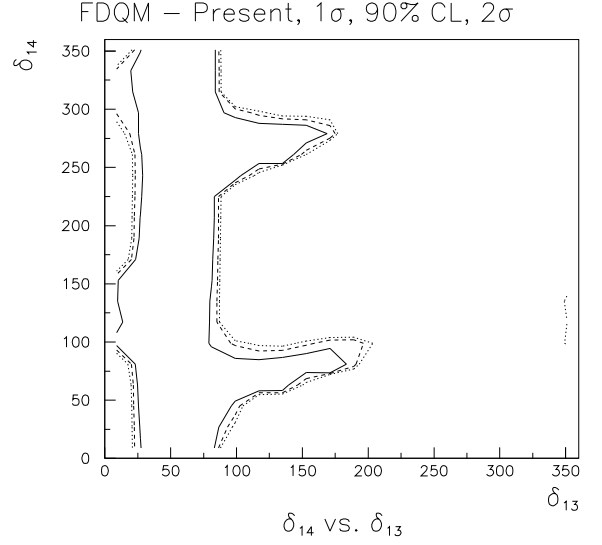}
\caption{\label{phases}
  Contour plot of $\delta_{14}$ vs. $\delta_{13}$ with contours at
  1$\sigma$, 90\% CL, and 2$\sigma$.}
\end{figure}

The cancellation in $V_{4s}$ to make it small requires $\delta_{24}
\approx \delta_{14} + \pi$.  Thus we can display the phases in a two
dimensional plot of $\delta_{14}$ vs. $\delta_{13}$, as in Fig.\ 
\ref{phases}.
When $\delta_{13}$ is in its SM range of $40^\circ$ ($\sin(\gamma) =
0.64$) to $70^\circ$ ($\sin(\gamma) = 0.94$) the SM terms can be
dominant and the small FCNC amplitudes allow each $\delta_{14}$
equally.  For certain values of $\delta_{14}$, near $80^\circ$ and
$270^\circ$, the new physics amplitudes can be dominant and
$\delta_{13}$ can be large, leading to the enlarged $(\rho,\eta)$
contours that can reach $\eta \approx 0$ and extend beyond to
$\delta_{13} \le 200^\circ$ at 2$\sigma$.

\subsection{FCNC Phase Structure}
Using the $V_{4s}$ cancellation structure with $s_{24} = s_{12}
s_{14}$ and $\delta_{24} = \delta_{14}+\pi$, we can rewrite the
$V_{4i}$ matrix elements in terms of just one phase in the leading
terms
\begin{eqnarray}
V_{4d} & \approx & -s_{14} e^{i \delta_{24}} \approx s_{14} 
e^{i \delta_{14}} 
 \\
V_{4b} & \approx & -s_{34} 
 \\
V_{4s} & \approx & (s_{24} - s_{12} s_{14}) 
e^{i\delta_{14}} + s_{34} s_{23}.
\end{eqnarray}
To leading order, it is clear that only the two new phases (of which
only one is effectively independent) are included in the $V_{4i}$, and
therefore in the $U_{ij}$ and in the FCNC amplitudes.  The SM phase
$\delta_{13}$ does not appear in the leading terms of $U_{ij}$.

The FCNC couplings, using the cancellation relations, are
\begin{eqnarray}
U_{ds} & = & -s_{14}\left[ (s_{24}-s_{12}s_{14})+
s_{34} s_{23} e^{-i\delta_{14}} \right],  \\
U_{sb} & = & s_{34} \left[ (s_{24}-s_{12}s_{14})e^{-i\delta_{14}}
+ s_{34} s_{23} \right], \\
U_{db} & = & s_{14} s_{34} e^{-i\delta_{14}}. 
\end{eqnarray}
We note that while $s_{14}$ and $s_{24}$ are non-zero, 
the cancellation in $V_{4s}$ and the ability of $s_{34}$ to vanish 
still allow all $U_{ij}$ to vanish.

\subsection{Variable Determination}
In general for the three complex matrix elements $V_{4i}$, one would
expect three magnitudes and three phases.  In determining these from
the $-U_{ij}=V^*_{4i}V_{4j}$ however, one overall phase would not
appear experimentally, due to the $V^*V$ structure of the $U_{ij}$.
So we can at best determine three magnitudes and two phases from the
$U_{ij}$.  This agrees with the three new angles and two new phases
introduced in the $4\times4$ unitary matrix where an extra phase has
been removed for the definition of the new $D^0$ down quark.  Whereas
the three $U_{ij}$ may seem to contain three real and imaginary parts
to be determined, they are not independent, since there is one
restriction between them, namely that the product
\begin{equation}
U_{ds}U_{sb}U_{bd} = |V_{4d}|^2 |V_{4s}|^2 |V_{4b}|^2 
\end{equation}
is real.  So again, we are left with three magnitudes and two phases
that can be determined by experiments involving the FCNC amplitudes,
which allows us to determine the three new angles and two new phases,
just from low energy experiments involving the $U_{ij}$.

With sufficient energy to produce one $D$ quark, the angles $s_{i4}$
can each be determined separately by the combined weak production of
$\bar{u}D$, $\bar{c}D$ or $\bar{t}D$ pairs, or from the similar decays
of the $D$ quarks.

The cancelation in $V_{4s}$ has related $s_{24} = \lambda s_{14}$ and
$\delta_{24} = \delta_{14} + \pi$.  Thus there are only effectively
two independent new angles and one new phase to be determined from the
five independent components of the $U_{ij}$, leading to an
overconstrained system.  Finding a consistent solution is then a test
of the FDQM.  Of course, if more variables are found to be needed, the
mixings to five or six down quarks would have to be considered.  The
present fits have found non-zero values for $s_{14}$ and its related
$s_{24}$.  Yet $s_{34}$ may still be small or vanish, and the one new
independent phase is still to be determined, although its
determination is coupled to that of the CKM $\delta_{13}$ phase.
 
\subsection{Unitarity Tests on the CKM Submatrix}
Contained in the $4\times4$ analysis are tests of the unitarity of the
$3\times3$ CKM submatrix contained in the $4\times 4$ FDQM mixing
matrix.  The FCNC couplings $U_{ds}$, $U_{sb}$, and $U_{db}$ measure
the deviations from orthogonality of the columns of the CKM submatrix,
in $d-s$, $s-b$, and $d-b$ projections, respectively.  Their sizes
will be discussed in subsection G under unitarity quadrangles.

Bounds on the size of the $|V_{4i}|$, i = 1, 2, 3, bounds the
deviation from unity of the sum of the squares of the three CKM
elements in each column
\begin{equation}
1 - (|V_{ui}|^2 + |V_{ci}|^2 + |V_{ti}|^2) = |V_{4i}|^2.
\end{equation}
Similarly, for the rows, the $|s_{i4}|^2$ measure the deviation from
unity for the sum of the squares of the CKM row elements.

For the $d$ column or $u$ row, since $|V_{4d}| \cong s_{14} \approx
0.035$ to 0.085, unitarity of the CKM three elements of the $d$ column
or $u$ row is off by
\begin{eqnarray}
0.0012 \le & |V_{4d}|^2 & \le 0.0072, ~~{\rm or} \\
0.5 \lambda^4 \le & |V_{4d}|^2 & \le 3 \lambda^4. 
\end{eqnarray}

For the $s$ column, since $|V_{4s}| \le 0.40 \times 10^{-3}$, the
deviation from unitarity of the CKM submatrix is bounded by
\begin{equation}
|V_{4s}|^2 \le 0.16 \times 10^{-6} = 0.6 \lambda^{10}.
\end{equation}

For the $b$ column or $t$ row, since $|V_{4b}|\cong s_{34} \le 0.020$,
the deviation from unitarity of the CKM submatrix is bounded by
\begin{equation}
|V_{4b}|^2 \le 0.00040 = 0.17 \lambda^4 \approx \lambda^5.
\end{equation}

For the $c$ row, since $s_{24} \cong s_{12} s_{14} = \lambda s_{14}$,
the deviation of the CKM from unitarity is a multiple of the $u$ row
result from $s_{14}$
\begin{equation}
0.5 \lambda^6 \le |s_{24}|^2 \le 3 \lambda^6.
\end{equation}

Finally, the deviation of $|V_{44}|^2$ from unity is an overall
measure of mixing to the fourth down quark
\begin{equation}
1 - |V_{44}|^2 = s_{14}^2 + s_{24}^2 + s_{34}^2.
\end{equation}
The right hand side is dominated by $s_{14}^2$ giving
\begin{equation}
|V_{44}|^2 = 1 - (0.5 \to 3) \lambda^4.
\end{equation}

\subsection{Unitarity Quadrangle Completion}
\subsubsection{$b-d$ Quadrange}
\begin{figure}%[ht]
%\centerline{\epsfbox{bwquad.eps}}
\includegraphics[width=8cm]{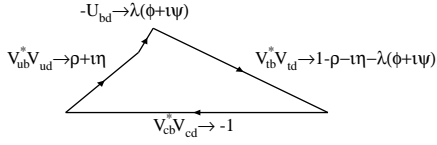}%
\caption{\label{quad} The $b-d$ unitarity quadrangle scaled by $A\lambda^3$, 
  with sides given as above.}
\end{figure}

The orthogonality relation between the $b$ and $d$ columns of the
$4\times4$ mixing matrix is
\begin{equation}
V^*_{ub}V_{ud}+V^*_{cb}V_{cd}+V^*_{tb}V_{td}-U^*_{db}=0.
\end{equation}
The fourth side of the $b-d$ unitarity quadrangle, scaled to make the
base of unit length is $U^*_{db}/|V^*_{cd}V_{cb}|$.  From Fig.\ \ref{Udb},
we see that the length of the FCNC quadrangle side is $\le 0.15$ in
the vertical or imaginary direction, and $\le 0.06$ in the horizontal
or real direction.
The sides of the $b-d$ unitarity quadrangle can be written in a
modified Wolfenstein form as
\begin{eqnarray}
V^*_{ub}V_{ud} &=& A\lambda^3(\rho+i\eta), \\
V^*_{cb}V_{cd} &=& -A\lambda^3, \\
V^*_{4b}V_{4d} &=& -U^*_{db} = - s_{14} s_{34} e ^{i\delta_{14}} \\
               &\equiv& A\lambda^4(\phi+i\psi), {\rm ~and}\\
V^*_{tb}V_{td} &=& A\lambda^3\left[1-\rho-i\eta-\lambda(\phi+i\psi)\right],
\end{eqnarray}
where we have introduced $\phi + i\psi$ into $-U^*_{db}$ with a
coefficient to make the scaled quadrangle $A$ independent.  An example
of the scaled $b-d$ quadrangle is shown in Fig.\ \ref{quad}.
We see that unitarity requires that the length of the FCNC coupling
side $-U^*_{db}$ has to be cancelled by another triangle side to close
the triangle, and that occurs in $V^*_{tb}V_{td}$ having an addition
to the SM formula.  The area of the $b-d$ unitarity quadrangle is
computed by adding the areas of three sub-triangles and a rectangle
\begin{equation}
{\rm Area}(b-d) = A^2\lambda^6\left[\eta+(1-\rho)\psi\right]/2.
\end{equation}
We note that if either or both $\eta$ and $\psi$ are non-zero,
$CP$ is violated, and the quadrangle has a non-zero area, analogous
to the SM unitarity triangle result.  However, as we will see below,
the area of the $b-d$ quadrangle is different from those of the
other unitarity quadrangles by the $\psi$ term above.

\subsubsection{$s-b$ Quadrangle}
\begin{figure}%[ht]
  %\epsfxsize = 5in 
%\centerline{\epsfbox{proj2prlog.eps}}
\includegraphics[width=8cm]{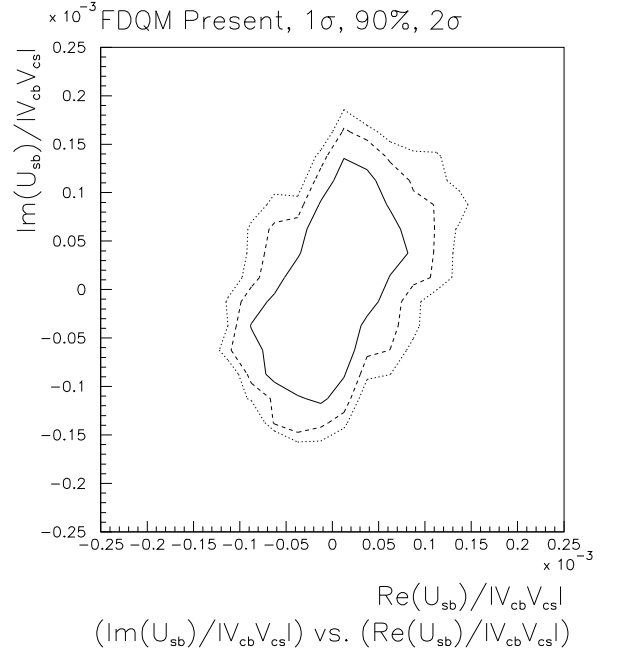}
\caption{\label{Usb} Contours for the complex FCNC coupling $U_{sb}$
  scaled by $|V^*_{cs}V_{cb}|$, which is the fourth side of the $s-b$
  unitarity quadrangle.}
\end{figure}

The unitarity orthogonality between the $s$ and $b$ columns for the 
$s-b$ quadrangle is
\begin{equation}
V^*_{us}V_{ub} + V^*_{cs}V_{cb} +  V^*_{ts}V_{tb} - U_{sb}=0.
\end{equation}
The first term is $A\lambda^4 (\rho-i\eta)$, the second term is
$A\lambda^2$, and the third term is $-A\lambda^2$, to leading order.
If we scale the base to unit length by dividing by $A\lambda^2$, then
the first term side is of order 0.02 in length.  From Fig.\ \ref{Usb},
the fourth side of scaled $U_{sb}$ is of order 0.0001, or 0.5\% of the
small third side of the triangle.  The enclosed angle is then the
same as in the SM, $\phi_s = \lambda^2 \eta$, and the triangle's or
quadrangle's area is $A^2\lambda^6\eta/2$.

\subsubsection{$d-s$ Quadrangle}
\begin{figure}%[ht]
%\epsfxsize = 5in
%\centerline{\epsfbox{proj1prlog.eps}}
\includegraphics[width=8cm]{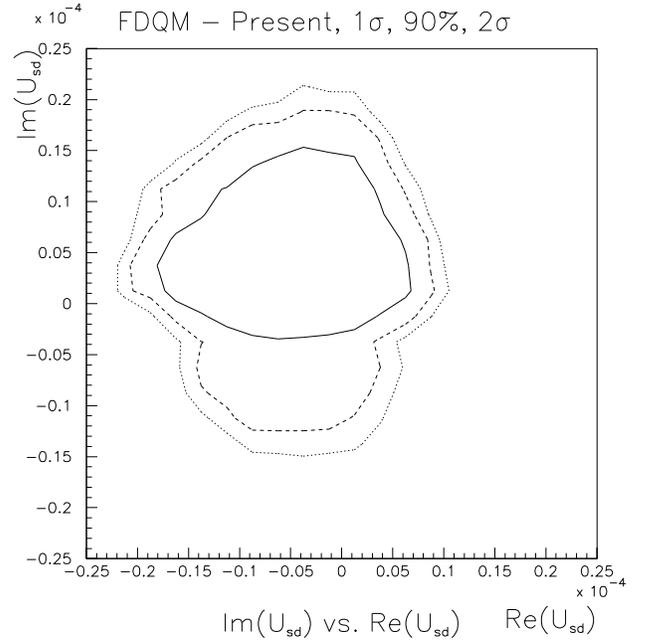}
\caption{\label{Uds} Contours for the complex FCNC coupling $U_{ds}$
which is the fourth side of the $d-s$ unitarity
quadrangle.}
\end{figure}

The orthogonality relation between the $d$ and $s$ columns is
\begin{equation}
V^*_{ud}V_{us} + V^*_{cd}V_{cs} + V^*_{td}V_{ts} - U_{ds} = 0.
\end{equation}
The largest sides of the $d-s$ unitarity quadrangle are of length
$\lambda$, being the first and second terms, and the third term is
$A^2\lambda^5(1-\rho+i\eta)= 0.0004(1-\rho+i\eta)$.  The fourth side
is the FCNC coupling $U_{sd}$, which is bounded in magnitude by
$2.5\times 10^{-5} = \lambda^7$, as seen in Fig.\ \ref{Uds}.  Thus the
FCNC fourth side is at most 6\% of the small third side.  The angle
subtended by the small third side is then essentially the same as that
by the third and fourth sides, being $\phi_d = A^2\lambda^4\eta$.  The
triangle's or quadrangle's area is also $A^2\lambda^6\eta/2$.

\subsection{The Sum Rule for the {\boldmath $CP$} violating 
{\boldmath $B$} Decay Asymmetry Angles} 
It was shown before \cite{Nir:1990hj} that
as long as the Penguin diagrams in the $B$ decays can be neglected,
that the sum of the $CP$ violating decay angles, even with new physics
contributions, is $\pi$ modulo $\pi$.  This can be seen from
Eqs.~(\ref{sinalpha}) and (\ref{sinbeta}) where in the sum of
$(2\alpha+2\beta)$, the $B_d$ mixing phase cancels out in general,
regardless of its source, and from Eqs.~(\ref{singamma}) and
(\ref{phis}), where in the sum of $(2\gamma+4\phi_s)$, the phase from
$B_s$ mixing cancels out.  The other tree amplitude decay phases in
these equations either cancel or sum to the phase of a product of
mixing matrix elements which becomes a product of absolute values
squared, with zero phase.  This leads to the $CP$ violating $B$ decay
angle sum rule \cite{Nir:1990hj}
\begin{equation}
\alpha + \beta + \gamma + 2\phi_s = \pi, \quad {\rm mod~} \pi.
\end{equation} 

%**************** Conclusions
\section{Conclusions for Iso-singlet Down Quark Models}
With much new data, it is still the case that FCNCs can contribute
significantly to $B_d -\bar{B_d}$ mixing and to $B_s-\bar{B}_s$
mixing, and give contributions with new phases.  In the FDQM, all
$\sin{(2\alpha)}$ are allowed.  In the $(\rho,\eta)$ plane, the FDQM
allows large regions for $\rho \le 0$ as opposed to the $\rho \ge 0$
regions in the SM, and in particular, those where $\eta$ goes to zero,
both with the present data and with the projected $\sin{(2\alpha)}$
values.  In new physics models then, the SM phase $\delta$, or $\eta$,
can be smaller, with the other phases causing much of the presently
observed $CP$ violation.  In the $(x_s,\sin{(\gamma)})$ plots in the
FDQM, all of $\sin(\gamma) \ge 0$ is allowed at present in contrast to
$\sin(\gamma) \ge 0.55$ in the SM, and with no approximately linear
relation as in the SM.  This will require combining experimental
results of $x_s$ and $\sin{(\gamma)}$, to find out if the results
correlate to the narrow linear region of the SM analysis.  The present
range for $x_s$ is from 16 to 48 at 2-$\sigma$ in the SM, and from 16
to 80 at 2-$\sigma$ in the four down quark model.  The $b-d$ unitarity
triangle, scaled to unit base length, has to be measured to an
accuracy of 0.15 or better to begin to limit a fourth side and to
verify the SM against the FDQM.

Each E$_6$ generation also contains an iso-singlet \cite{Hewett:1989xc}
or sterile neutrino, which may provide a connection between the quark
and lepton searches for new physics in terms of establishing new
particle representations.

The mass of the lightest singlet down quark in E$_6$ could be roughly
related to the mixing angle by
\begin{equation}
\theta_{34}^2 \simeq  m_b/m_D, \quad {\rm and~ with} ~~
|V_{4b}| \simeq \theta_{34} \leq 0.02 
\end{equation}
from combined fits, that gives
\begin{equation}
m_D \geq 2500 \times m_b = 11 \ {\rm TeV} .  
\end{equation}
Using the single $R_b$ 90\% CL limit of $|\theta_{34}| \leq 0.035$,
which is not as strong as the combined fits, gives $m_D \geq 4$ TeV.
The previous analysis \cite{Silverman:1998uj} gave a lower limit of 1.2
TeV.

\begin{acknowledgments}
  This research was supported in part by the U.S. Department of Energy
  under Contract No. DE-FG0391ER40679.  We acknowledge the hospitality
  of SLAC.  We thank Herng Tony Yao, Sheldon Stone, and David Kirkby
  for discussions.
\end{acknowledgments}

%\bibliography{refs01}
%\bibliographystyle{apsrev}

\printfigures
\end{document}